\documentclass[aps,pra,preprint,groupedaddress]{revtex4}
\usepackage{amssymb,bm}
\usepackage[dvips]{graphicx}

\begin{document}
\bibliographystyle{apsrev}

\title{Relativistic Coulomb Green's function in $d$-dimensions.}
\author{R.N.~Lee}
\author{A.I.~Milstein  }
\author{I.S.~Terekhov}
\affiliation{Budker Institute of Nuclear Physics\\
and Novosibirsk State University,\\ 630090 Novosibirsk, Russia }

\date{\today}

\begin{abstract}
Using the operator method, the Green's functions of the Dirac and Klein-Gordon equations in
the Coulomb potential $-Z\alpha/r$ are derived for the arbitrary space dimensionality $d$.
Nonrelativistic and quasiclassical asymptotics of these Green's functions are considered in
detail.
\end{abstract}


\maketitle

\section{Introduction}
When calculating the amplitudes and probabilities of QED processes in the field of heavy
atoms, one should be aware that the parameter $Z\alpha$  ($Z$ is the atomic charge number
and $\alpha$ is the fine-structure constant) is not small in this case. The effect of higher
orders in $Z\alpha$ can change the Born result by several times. Therefore, it is often
required to calculate the probabilities of QED processes in such a strong field exactly in
$Z\alpha$. The most convenient way to perform this calculation is the use of  the exact
Green's functions of the Dirac equation (or Klein-Gordon equation) for a charged particle in
the field (Furry representation). Therefore, derivation of the Green's functions for specific field
configurations is very important for applications. For the case of the Coulomb potential, a
convenient integral representation of the Green's function $G(\bm r,\bm r' |\varepsilon)$ was
derived in Ref. \cite{MS82} using the $O(2,1)$ algebra. The representation obtained is valid in
the whole complex plane of the energy $\varepsilon$  and does not contain contour integrals.
Another integral representation for  the Green's function in a Coulomb field was derived in
Ref.\cite{Host64} by using the explicit form of the expansion of $G(\bm r,\bm r' |\varepsilon)$
with respect to the eigenfunctions of the corresponding wave equation. The representation of
the Green's function obtained in Ref.\cite{Host64} contains a contour integral, which
complicates its use in applications.

In the calculation of the loop diagrams, it is often required to regularize the divergent integrals.
One of the most convenient methods of the regularization is the dimensional regularization. In
order to use the dimensional regularization within the approach based on the Furry
representation, it is necessary to derive the exact Green's function in the Coulomb field in
arbitrary, not necessarily integer, space dimensionality $d$ (the space-time dimensionality is
$d+1$). In the present paper, we solve this problem by generalizing the Green's function,
obtained in Ref. \cite{MS82} for $d=3$, to arbitrary $d$. Our derivation closely follows the
path of derivation in Ref. \cite{MS82}. In contrast to the conventional approach, the operator
method used in Ref. \cite{MS82} and in the present paper does not require the knowledge of
the explicit form of the wave functions, which is difficult to define for non-integer $d$. In order
to fix unambiguously the explicit form of the Green's function for arbitrary $d$, we use only the
commutative and anticommutative relations for the operators and $\gamma$-matrices .

\section{Calculation of the Green's function}

Following Ref. \cite{MS82}, we represent the Green's function in the Coulomb potential
$U(r)=-Z\alpha/r$ (the system of units $\hbar=c=1$ is used),
\begin{equation}
G(\bm r,\bm r' |\varepsilon)=\frac{1}{\hat{\cal P}-m+i0}\delta(\bm r-\bm r')\,,
\quad \hat{\cal P}=\gamma^0(\varepsilon+Z\alpha/r)-\bm\gamma\bm p\,,
\end{equation}
as follows
\begin{eqnarray}\label{GD}
G(\bm r,\bm r' |\varepsilon)&=&(\hat{\cal P}+m)D(\bm r,\bm r' |\varepsilon)\,\nonumber\\
D(\bm r,\bm r' |\varepsilon)&=&-i\int_0^\infty ds \exp\left\{2iZ\alpha\varepsilon s -is\left[rp_r^2+\kappa^2r+\frac{K}{r}\right]\right\}
\frac{\delta(r-r')}{r^{d-2}}\delta(\bm n-\bm n')\,,\nonumber\\
&&\kappa=\sqrt{m^2-\varepsilon^2}\,,\quad p_r=-\frac{i}{r^{(d-1)/2}}\frac{\partial}{\partial r}r^{(d-1)/2}\,,
\quad \bm n=\bm r/r\,,\quad \bm n'=\bm r'/r'\,,\nonumber\\
&&K=\bm l^2-iZ\alpha \bm\alpha\bm n-(Z\alpha)^2+\frac{1}{4}(d-1)(d-3)\,,\quad \bm\alpha=\gamma^0\bm\gamma\,.
\end{eqnarray}
Here $-\bm l^2$ is the angular part of Laplacian determined by
\begin{equation}
\Delta=\frac1{ r^{d-1}}\partial_r r^{d-1}\partial_r-\frac1{r^2}\bm l^2\,.
\end{equation}
and $\gamma$-matrices obey the usual relation $\gamma^\mu \gamma^\nu+\gamma^\nu
\gamma^\mu=g^{\mu\nu}$.

Then we represent the angular part of the $\delta$-function as
\begin{equation}
\delta(\bm n-\bm n')=\sum_\lambda P_\lambda(\bm n,\bm n')\,,
\end{equation}
where the  projection operators $P_\lambda(\bm n,\bm n')$ satisfy the relations
\begin{eqnarray}
&&K\,P_\lambda(\bm n,\bm n')=\lambda(\lambda+1)P_\lambda(\bm n,\bm n')\,,
\nonumber\\
\label{eqP}&&\int d\bm n' P_\lambda(\bm n,\bm n')P_{\lambda'}(\bm n',\bm n'')=\delta_{\lambda\lambda'}P_\lambda(\bm n,\bm n'')\,.
\end{eqnarray}
Since the operator $K$ contains only one matrix operator $\bm\alpha\bm n$, the matrix
structure of the projection operator $P_\lambda(\bm n,\bm n')$ is given by the linear
combination of the unit matrix $I$ and matrices $\bm\alpha\bm n$, $\bm\alpha\bm n'$, and
$(\bm\alpha\bm n)(\bm\alpha\bm n')$. All other matrices, such as $(\bm\alpha\bm
n)(\bm\alpha\bm n')(\bm\alpha\bm n)$, can be reduced to the four above mentioned matrices
using the anticommutative relations. Taking this property into account, we search the
projection operators $P_\lambda(\bm n,\bm n')$ in the form
\begin{eqnarray}
P_\lambda(\bm n,\bm n')&=&a_{1}\Lambda_+(\bm n)\Lambda_+(\bm n')+a_{2}\Lambda_+(\bm n)\Lambda_-(\bm n')
\nonumber\\
&&+a_{3}\Lambda_-(\bm n)\Lambda_+(\bm n')+a_{4}\Lambda_-(\bm n)\Lambda_-(\bm n')\,,\nonumber\\
&&\Lambda_\pm(\bm n)=\frac{1}{2}(1\pm \bm\alpha\bm n)\,,
\end{eqnarray}
where $a_{i}$ are some functions of $x=\bm n\bm n'$. From Eqs. (\ref{eqP}), we obtain
\begin{eqnarray}\label{ai}
&&a_1=\beta (\lambda+iZ\alpha)B_n(x)\,,\quad a_2=a_3=\beta\left(n+\nu+1/2\right)A_n(x)\,,
\quad a_4=\beta (\lambda-iZ\alpha)B_n(x)\,,\nonumber\\
&&\lambda=\pm\gamma\,,\quad \gamma=\sqrt{\left(n+\nu+1/2\right)^2-(Z\alpha)^2}\,,
\quad \beta=\frac{\Gamma(\nu+1)}{2\lambda\pi^{\nu+1}}\,,
\nonumber\\
&&A_n(x)=\frac{1}{2\nu}\frac{\partial}{\partial x}[C_{n+1}^\nu(x)+C_{n}^\nu(x)]\,,\quad
B_n(x)=\frac{1}{2\nu}\,\frac{\partial}{\partial x}[C_{n+1}^\nu(x)-C_{n}^\nu(x)]\,,\nonumber\\
&&\nu=\frac{d}{2}-1\,,
\end{eqnarray}
where $C_{n}^\nu(x)$ is the Gegenbauer polynomial, and $n=0,1,2,\ldots$ is integer number.
This integer number appears from the  requirement that the functions $a_i$ have no
singularities at $x=1$. The result (\ref{ai}) for $a_i$ was obtained with the help of the identity
\begin{eqnarray}
&&\int(1+\bm n\bm n'+\bm n\bm n''+\bm n'\bm n'')B_n(\bm n\bm n')
B_n(\bm n'\bm n'')\,d\bm n'=\Omega_d(1+\bm n\bm n'')B_n(\bm n\bm n'')\,,\nonumber\\
&&\Omega_d=\int d\bm n=\frac{2\pi^{d/2}}{\Gamma(d/2)}=\frac{2\pi^{\nu+1}}{\Gamma(\nu+1)}\,,
\end{eqnarray}

 Finally we obtain for projection operator
 \begin{eqnarray}\label{Pfinal}
P_\lambda(\bm n,\bm n')&=&\frac{\beta}{2}\Bigg\{\Big[\lambda [1+(\bm\alpha\bm n)(\bm\alpha\bm n')]
+iZ\alpha(\bm\alpha\bm n+\bm\alpha\bm n')\Big]B_n(x)
\nonumber\\
&&+\left(n+\nu+1/2\right)[1-(\bm\alpha\bm n)(\bm\alpha\bm n')]A_n(x)\Bigg\}\,.
\end{eqnarray}
For $d=3$, this projection operator coincides with that found in Ref.\cite{MG58}.

Note that the functions $A_n(x)$ and $B_n(x)$ have non-singular limit at $\nu\to 0$ (or $d\to
2$) ,
$$\lim_{\nu\to 0}A_n(x)=\frac{\sin((n+1)\phi)+\sin(n\phi)}{\sin\phi}\,,\quad
 \lim_{\nu\to 0}B_n(x)=\frac{\sin((n+1)\phi)-\sin(n\phi)}{\sin\phi}\,,$$
 with $\phi=\arccos x$.

In order to complete the calculation of the function $D(\bm r,\bm r' |\varepsilon)$, Eq.
(\ref{GD}), it is necessary to find the result of the action of the operator
$\exp\left\{-is\left[rp_r^2+\kappa^2r+\lambda(\lambda+1)/r\right]\right\}$ on the function
$\delta(r-r')/r^{2\nu}$. This can be done  exactly in the same way as in  Ref.\cite{MS82}. The
method of Ref.\cite{MS82} is based on the commutator relations of the operators
$T_1=\frac{1}{2}[rp_r^2+\lambda(\lambda+1)/r]$, $T_2=rp_r$, and $T_3=r$ which coincide
with those of the $O(2,1)$ algebra generators (some other examples of applying the $O(2,1)$
algebra in a Coulomb field can be found in Refs.\cite{Nambu67,DR1970}). The only difference
between the case of arbitrary $d$ and $d=3$ is the value of the parameter $\delta$ in the
equation $T_1r^\delta=0$. For arbitrary $d$, we have
\begin{equation}\label{delta}
\delta=\lambda+\frac{3-d}{2}\quad \mbox{at}\, \lambda>0\,,\quad \delta=|\lambda|+\frac{1-d}{2}\quad \mbox{at}\, \lambda<0\,.
\end{equation}
The final result for the function $D(\bm r,\bm r' |\varepsilon)$ in Eq.(\ref{GD}) reads
\begin{eqnarray}\label{Dfinal}
D(\bm r,\bm r' |\varepsilon)&=&-\frac{i\Gamma\left(\nu+1\right)}{2\pi^{\nu+1}(rr')^{\nu+1/2}}\sum_{n=0}^\infty
\int_0^\infty ds \exp[2iZ\alpha\varepsilon s+i\kappa (r+r')\cot (\kappa s)-i\pi\gamma]\nonumber\\
&&\times\Bigg\{\frac{y}{2}J'_{2\gamma}(y)[1+(\bm\alpha\bm n)(\bm\alpha\bm n')]B_n(x)+iZ\alpha J_{2\gamma}(y) (\bm\alpha\bm n+\bm\alpha\bm n')B_n(x)\nonumber\\
&&+\left(n+\nu+1/2\right)J_{2\gamma}(y)[1-(\bm\alpha\bm n)(\bm\alpha\bm n')]A_n(x)\Bigg\}\,,
\quad y=\frac{2\kappa\sqrt{rr'}}{\sin(\kappa s)}\,,
\end{eqnarray}
where $J_{2\gamma}(y)$ is the Bessel function, $A_n(x)$, $B_n(x)$, $\nu$ and $\gamma$
are defined in Eq.(\ref{ai}). The corresponding result for the Coulomb Green's function of the
Dirac equation in $d$ space dimension has the form
\begin{eqnarray}\label{Gfinal}
&&G(\bm r,\bm r' |\varepsilon)=-\frac{i\Gamma\left(\nu+1\right)}{2\pi^{\nu+1}(rr')^{\nu+1/2}}\sum_{n=0}^\infty
\int\limits_0^\infty ds \exp[2iZ\alpha\varepsilon s+i\kappa (r+r')\cot (\kappa s)-i\pi\gamma]\,{\cal T}\nonumber\\
&&{\cal T}=[1+(\bm\alpha\bm n)(\bm\alpha\bm n')]\Big[\frac{y}{2}J'_{2\gamma}(y)(\gamma^0\varepsilon+m)
-iZ\alpha J_{2\gamma}(y)\gamma^0\kappa\cot(\kappa s)\Big]B_n(x)\nonumber\\
&&+\Big[[1-(\bm\alpha\bm n)(\bm\alpha\bm n')](\gamma^0\varepsilon+m)
-\kappa\cot(\kappa s)(\bm\gamma\bm n-\bm\gamma\bm n')\Big]
J_{2\gamma}(y)\left(n+\nu+1/2\right)A_n(x)\nonumber\\
&&+\left[\frac{i\kappa^2(r-r')}{2\sin^2(\kappa s)}+imZ\alpha\gamma^0
\right] (\bm\gamma\bm n+\bm\gamma\bm n')J_{2\gamma}(y)B_n(x)\, .
\end{eqnarray}
For $d=3$ this result coincides with the corresponding result of Ref.\cite{MS82}. The function
$G(\bm r,\bm r'|\,\varepsilon)$ has, in the complex plane $\varepsilon$, cuts along the real axis
from $-\infty$ to $-m$ and from $m$ to $\infty$, which correspond to the continuous spectrum,
and has also simple poles in the interval $(0,m)$ for an attractive field and in the interval $(-m,
0)$ for a repulsive field. The integral representation (\ref{Gfinal}) is valid for any $\varepsilon$
which belongs to the domain $\mbox{Re}\,\varepsilon <0,\, \mbox{Im}\,\varepsilon <0$ or
$\mbox{Re}\,\varepsilon>0,\, \mbox{Im}\,\varepsilon >0$. If $\mbox{Re}\,\varepsilon <0,\,
\mbox{Im}\,\varepsilon>0$ or $\mbox{Re}\,\varepsilon >0,\, \mbox{Im}\,\varepsilon <0$, then it
is necessary to perform the integration over $s$ in Eq.(\ref{Gfinal}) from zero to $-\infty$.

For real $\varepsilon$ in the  interval $-m <\varepsilon <m$ we obtain (cf. Ref.\cite{MS82})
\begin{eqnarray}\label{Gfinald}
&&G(\bm r,\bm r' |\varepsilon)=\frac{\Gamma\left(\nu+1\right)}{4\kappa\sin[\pi (Z\alpha\varepsilon/\kappa-\gamma)]
\pi^{\nu+1}(rr')^{\nu+1/2}}\sum_{n=0}^\infty
\int\limits_{-\pi/2}^{\pi/2} ds \nonumber\\
&&\times\exp\left[-2iZ\alpha\varepsilon s/\kappa+i\kappa (r+r')\tan s\right]\,{\cal T}\nonumber\\
&&{\cal T}=[1+(\bm\alpha\bm n)(\bm\alpha\bm n')]\Big[\frac{v}{2}J'_{2\gamma}(v)(\gamma^0\varepsilon+m)
-iZ\alpha J_{2\gamma}(v)\gamma^0\kappa\tan s\Big]B_n(x)\nonumber\\
&&+\Big[[1-(\bm\alpha\bm n)(\bm\alpha\bm n')](\gamma^0\varepsilon+m)
-\kappa\tan s(\bm\gamma\bm n-\bm\gamma\bm n')\Big]
J_{2\gamma}(v)\left(n+\nu+1/2\right)A_n(x)\nonumber\\
&&+\left[\frac{i\kappa^2(r-r')}{2\cos^2 s}+imZ\alpha\gamma^0
\right] (\bm\gamma\bm n+\bm\gamma\bm n')J_{2\gamma}(v)B_n(x)\, , \quad v=\frac{2\kappa\sqrt{rr'}}{\cos s}\, .
\end{eqnarray}
The denominator in Eq. (\ref{Gfinald}) is zero at points $Z\alpha\varepsilon/\kappa -\gamma =k$ for any integer $k$. However, the integral over $s$ also
vanishes for negative $k$ at these points (see Ref.\cite{MS82}) so that the expression(\ref{Gfinald}) has poles only for
 $k =0,\, 1,\,2\ldots$. Taking into account that  $\gamma$  is positive,
we find that  the simple poles corresponding to the discrete spectrum  are at the points
\begin{equation}
\varepsilon=\frac{m\,\mbox{sign}Z}{\sqrt{1+ \left(\frac{Z\alpha}{k+\gamma}\right)^2}}\, .
\end{equation}
The maximal value of $Z$ when all the results
obtained are applicable is determined by the relation $(Z\alpha)_{max}=(d-1)/2$ (see the
definition of $\gamma$ in Eq.(\ref{ai})).

For completeness, we also present the final result for the  Coulomb Green's function of the
Klein-Gordon equation,
\begin{eqnarray}\label{KGGfinal}
&&G_0(\bm r,\bm r' |\varepsilon)=-\frac{\Gamma\left(\nu+1\right)}{2\pi^{\nu+1}(rr')^{\nu}}\sum_{n=0}^\infty
\frac{n+\nu}{\nu}\,C_n^\nu(x)\nonumber\\
&&\times\int\limits_0^\infty \frac{\kappa\, ds}{\sin(\kappa s)} \exp[2iZ\alpha\varepsilon s+i\kappa (r+r')\cot (\kappa s)-i\pi\mu]J_{2\mu}(y)\, ,\nonumber\\
&&\mu=\sqrt{\left(n+\nu\right)^2-(Z\alpha)^2}\,.
\end{eqnarray}
Note that there is no singularity in this formula at $d=2$ because
$$
\lim_{\nu\to 0}\frac{n+\nu}{\nu}\,C_n^\nu(x)=\cos(n\phi)\,.
$$
\section{Asymptotics}
Let us derive the Coulomb Green's function $G_{nr}(\bm r,\bm r' |E)$ of the Schr\"o{}dinger
equation for $d$ space dimensions. In order to do this we calculate the nonrelativistic
asymptotics of the Coulomb Green's function of the Klein-Gordon equation valid at $|E|\ll m$
 and $Z\alpha\ll 1$, where $E=\varepsilon-m$. Neglecting  $(Z\alpha)^2$ in $\mu$, using the
formula of summation (cf. \cite{Host64}),
\begin{eqnarray}\label{sum1}
S_0&=&\sum_{n=0}^\infty(-1)^n\frac{\nu+n}{\nu}C_n^\nu(x)J_{2(n+\nu)}(y)\nonumber\\
&=&\frac{\sqrt{\pi}\,y^{2\nu}J_{\nu-1/2}(w)}{2^{3\nu+1/2}\Gamma(\nu+1)w^{\nu-1/2}}\,,
\quad w=y\sqrt{\frac{1+x}{2}}
\end{eqnarray}
and multiplying by $2m$, we obtain
\begin{eqnarray}\label{Gnr}
&&G_{nr}(\bm r,\bm r' |E)=-\frac{m}{(2\pi)^{\nu+1/2}}\int\limits_0^\infty  ds \left(\frac{\kappa}{\sin(\kappa s)}\right)^{2\nu+1}
\nonumber\\
&&\times
 \exp[2iZ\alpha m s+i\kappa (r+r')\cot (\kappa s)-i\pi\nu]\frac{J_{\nu-1/2}(w)}{w^{\nu-1/2}}\,,\nonumber\\
&&\kappa=\sqrt{-2m E}\,.
\end{eqnarray}
This formula is in agreement with the corresponding result of Ref.\cite{Hostler70}.

At high energies and small scattering angles of the particles, the characteristic angular
momenta  are large and the quasiclassical approximation is applicable. The quasiclassical
Green's function  of the Dirac equation in a Coulomb potential for $d=3$ was first derived in Refs.
\cite{MS83,MS83A}. Another representation of this function  was obtained in Refs.
\cite{LMS97,LMS98}. The quasiclassical Green's function for arbitrary spherically symmetric
localized potential was found in Refs. \cite{LM95,LM95A}. In Ref.\cite{LMS2000}, the
quasiclassical Green's function for arbitrary  localized potential was found with the
next-to-leading quasiclassical correction taken into account. In Ref.\cite{LMS2000}, a
spherical symmetry of the potential was not required. Let us consider the quasiclassical
Green's function of the Dirac equation in a Coulomb potential for arbitrary space dimension
$d$. In this case $\varepsilon\gg m$ and $1+x\ll 1$ so that the main contribution to the sum
over $n$ in Eq.(\ref{Gfinal}) is given by $n\gg1$. Thus we can neglect the term $(Z\alpha)^2$
in $\gamma$, Eq.(\ref{ai}), and perform   summation over $n$ analytically. We need to
calculate two sums,
\begin{eqnarray}
S_A&=&\sum_{n=0}^\infty(-1)^n (\nu+n+1/2)\,A_n(x)J_{2(n+\nu+1/2)}(y)\,,\nonumber\\
S_B&=&\sum_{n=0}^\infty(-1)^n B_n(x)J_{2(n+\nu+1/2)}(y)\,,
\end{eqnarray}
where the functions $A_n(x)$ and $B_n(x)$ are defined in Eq.(\ref{ai}). Using the recurrent relations for
the Bessel functions and the Gegenbauer polynomials, it is easy to show that
\begin{eqnarray}
S_A&=&(1+x)\frac{\partial}{\partial x}S_B+(\nu+1/2) S_B\,,\quad
S_B=-\frac{2}{y}\frac{\partial}{\partial x}S_0\,,
\end{eqnarray}
so that
\begin{eqnarray}\label{sum2}
S_A&=&=\frac{\sqrt{\pi}\,y^{2\nu+1}J_{\nu-1/2}(w)}{2^{3\nu+5/2}\Gamma(\nu+1)w^{\nu-1/2}}\,,\quad
S_B=\frac{\sqrt{\pi}\,y^{2\nu+1}J_{\nu+1/2}(w)}{2^{3\nu+3/2}\Gamma(\nu+1)w^{\nu+1/2}}\,.
\end{eqnarray}
Substituting these results in Eq.(\ref{Gfinal}), we arrive at the final expression to the
quasiclassical Green's function
\begin{eqnarray}\label{Gqcl}
&&G_{qc}(\bm r,\bm r' |\varepsilon)=-\frac{1}{2^{\nu+3/2}\pi^{\nu+1/2}}
\int\limits_0^\infty \frac{ds}{u^{\nu-1/2}}\left(\frac{p}{\sinh(ps)}\right)^{2\nu+1}\nonumber\\
&&\times \exp[2iZ\alpha\varepsilon s+ip (r+r')\coth (ps)-i\pi\nu]\,{\cal M}\nonumber\\
&&{\cal M}=J_{\nu-1/2}(u)\left[\gamma^0\varepsilon+m-\frac{p}{2}\coth(ps)(\bm\gamma\bm n-\bm\gamma\bm n')\right]
+i\frac{J_{\nu+1/2}(u)}{u}
\nonumber\\
&&\times\Bigg\{\left[\frac{p^2(r-r')}{2\sinh^2(p s)}+mZ\alpha\gamma^0
\right](\bm\gamma\bm n+\bm\gamma\bm n')-Z\alpha\gamma^0p\coth(p s)[1+(\bm\alpha\bm n)(\bm\alpha\bm n')]\Bigg\}
\, ,\nonumber\\
&&u=\frac{p\sqrt{2rr'(1+x)}}{\sinh(ps)}\,,
\end{eqnarray}
where $p=\sqrt{\varepsilon^2-m^2}=i\kappa$. For $d=3$ the result (\ref{Gqcl}) is in agreement
with that obtained in Refs.\cite{MS83,MS83A}. The term $(Z\alpha)^2$ in $\gamma$,
Eq.(\ref{ai}), can be also neglected in the nonrelativistic approximation when $Z\alpha\ll1$,
$p\ll m$, and $Z\alpha m/p$ is fixed. In this case we immediately obtain from Eq.(\ref{Gqcl})
that the nonrelativistic approximation for the Green's function of the Dirac equation reads
\begin{equation}
G(\bm r,\bm r' |m+E)=\frac{\gamma^0+1}{2}G_{nr}(\bm r,\bm r' |E)\,,
\end{equation}
where $G_{nr}(\bm r,\bm r' |\varepsilon)$ is defined in Eq.(\ref{Gnr}).

To summarize,  we have calculated in $d$ space dimensions  the Green's functions of the
Dirac, Eq.(\ref{Gfinal}), and Klein-Gordon, Eq.(\ref{KGGfinal}), equations in the Coulomb field. Nonrelativistic and quasiclassical
limiting cases of these Green's functions are considered in detail. The results obtained can be
applied for calculation of various QED amplitudes in the strong Coulomb field with the use of
dimensional regularization.

This work was supported in part by the RFBR Grant No 09-02-00024 and the Grant 14.740.11.0082 of Federal Program "Personnel of Innovational Russia".  I.S.T. was also supported by the  "Dynasty" foundation.

\end{document}